\documentclass{aa}
\usepackage{graphics,latexsym,amssymb,times,psfig}

\def\gsim{ \lower .75ex \hbox{$\sim$} \llap{\raise .27ex \hbox{$>$}} } 
\def\lsim{ \lower .75ex\hbox{$\sim$} \llap{\raise .27ex \hbox{$<$}} } 

\begin{document}

\title{Clumps in large scale relativistic jets.}


\author{Fabrizio Tavecchio \inst{1}, Gabriele Ghisellini \inst{1} and 
Annalisa Celotti, \inst{2}}
\offprints{F. Tavecchio; tavecchio@merate.mi.astro.it}
\institute{INAF, Osserv. Astron. di Brera, via Bianchi 46, I--23807 Merate, Italy;
\and SISSA/ISAS, via Beirut 2-4, I--34014 Trieste, Italy. }

\date{Received 2002}
 
\titlerunning{Tomography of jets}
\authorrunning{Tavecchio, Ghisellini \& Celotti}

\abstract{ The relatively intense X--ray emission from large scale
(tens to hundreds kpc) jets discovered with Chandra likely implies that jets
(at least in powerful quasars) are still relativistic at that
distances from the active nucleus.  In this case the emission is due to
Compton scattering off seed photons provided by the Cosmic Microwave
Background, and this on one hand permits to have magnetic fields close
to equipartition with the emitting particles, and on the other hand minimizes
the requirements about the total power carried by the jet.
The emission comes from compact (kpc scale) knots, and we here
investigate what we can predict about the possible emission between
the bright knots.  This is motivated by the fact that bulk
relativistic motion makes Compton scattering off the CMB photons
efficient even when electrons are cold or mildly relativistic in the
comoving frame. This implies relatively long cooling times, dominated
by adiabatic losses.  Therefore the relativistically moving plasma can
emit, by Compton scattering the microwave seed photons, for a long
time.  We discuss how the existing radio--to--X--ray observations of
large scale jets already pose strong constraints on the structure and
dynamics of knots and we present a scenario that can satisfactorily
reproduce the observed phenomenology of the jet in 3C273.  In
this scenario the kiloparsec--scale knots visible with HST, Chandra
and VLA are composed of several smaller sub--units, accounting for the
fast decrease of the flux outside the large knot. Substructure in
the X-ray- emitting knots can also explain the month--year variability
timescale reported for the large scale jet in M87.
}
\maketitle

\section{Introduction}

Thanks to Chandra observations, we now know that the jets of
extragalactic radio sources have X--ray knots which are tens to hundreds of 
kpc away from the nucleus.
This seems to be true both for powerful flat radio spectrum  
sources (FSRQ) with a jet probably aligned with the line of sight
(Chartas et al. 2000; 
Pesce et al. 2001;
Sambruna et al. 2002;
Sambruna et al. 2001;
Schwartz 2002;
Schwartz et al. 2000;
Siemiginowska et al. 2002;
Tavecchio et al. 2000),
and also for closer radio--galaxies whose jets are observed at large 
viewing angles (e.g. 
Wilson \& Yang, 2002;
Kraft et al. 2002; 
Hardcastle et al., 2002; 
Hardcastle et al. 2001; 
Wilson et al. 2001).

In the powerful FSRQ the X--ray luminosity often exceeds the optical and
radio power emitted in the same knots.
Thermal and synchrotron self Compton models have difficulties
in explaining the data (see e.g. 
Chartas et al. 2000; 
Schwartz et al. 2000; 
Siemiginowska et al. 2002; 
Harris \&  Krawczynski 2002), 
while interpreting the X--rays as due to inverse Compton radiation off the cosmic 
microwave background (CMB), and the radio--optical emission as due to synchrotron, 
can well explain the observations and at the same time minimize the energy 
requirements (Celotti, Ghisellini \& Chiaberge 2001;
Tavecchio et al. 2000; Ghisellini \& Celotti 2001).
This however requires the jet to be highly relativistic even at the largest
scales: in the case of PKS 0637--712 the bulk Lorentz factor is of the order
of 10--15 at hundreds of kpc away from the core.

In this case the CMB energy density is seen boosted in the comoving
frame of the jet, by a factor $\Gamma^2$, causing it to dominate over
the synchrotron and the magnetic field energy densities existing
locally.

Another consequence of this scenario is that it is possible to
constrain the low energy cut off of the emitting particle distribution,
$\gamma_{\rm min} m_e c^2$.
Taking PKS 0637--752 as an example, we 
require $\gamma_{\rm min}<30$ in order that the inverse Compton spectrum
starts before the X--ray band (assuming $\Gamma=14$).
We also require $\gamma_{\rm min}>10$ to avoid the inverse Compton
process overproducing the observed optical emission.
This is true independent of whether the optical emission  belongs 
to the initial part of the IC spectrum or to the high energy 
tail of the  synchrotron flux. 
However, a synchrotron origin of the optical flux corresponds
to emission of ultrarelativistic electrons of relatively short cooling 
time, which nicely corresponds (albeit within the large uncertainties) 
to the dimensions of the optical knots as observed by HST 
(Schwartz et al. 2000; Celotti, Ghisellini \& Chiaberge 2001).

In this paper we consider sources observed at small viewing angles,
that can emit beamed X--rays by their plasma moving at bulk
relativistic speeds.  Since we see bright knots, we know that there is
some process accelerating particles at very high energies: to produce
optical radiation by synchrotron emission with reasonable values of
the magnetic field, we require electrons of TeV energies.  At these
scales, the cooling time of the electrons is relatively long for all
but the highest energy particles, and it is therefore of interest to
study what is the predicted spectral energy distribution (SED) {\it
outside} the bright knots, taking into account that the energy density
(in the comoving frame) of the CMB is constant for a non--decelerating
jet, and therefore the inverse Compton scattering of these photons is
always important.  On the other hand it is reasonable to assume that
the magnetic field becomes smaller for increasing distances from the
nucleus, causing the synchrotron luminosity to decrease even if the
emitting particles do not cool.
In this respect, it is intriguing that we may be able to detect
inverse-Compton radiation produced by particles that are only mildy
relativistic or even sub-relativistic in the comoving frame but have
substantial bulk motions. Is this ``bulk Compton" luminosity
observable by existing or future instruments, such as ALMA and NGST?

Furthermore, we will discuss observations which already exist, and
pose an interesting problem: the X--ray flux, outside the bright
knots, is dimming as fast as the optical and the radio fluxes, despite
the fact that it should be produced by low energy electrons, which do
not cool radiatively.  Adiabatic losses are required, but this of
course implies that the emitting region expands, and by calculating
how much expansion is required we will reach interesting conclusions
about the geometry of the acceleration sites.

In order to attack these problems we will first study how the emitting
particle distribution evolves in time.  In the presence of bulk
motion, the time evolution translates to a radial profile from the
acceleration site.  However, this correspondence is not unique, since
it depends on the nature of the bright knots we see: we do not know
yet if these correspond to {\it standing} or {\it moving} features.
In the former case relativistic particles originating in a shock move
out of it forming a trail; in the latter case there is no connection
between the bright knots (which are solidly moving together with the
emitting particles) and the interknot regions.

The main result of our analysis is that if the knot has a homogeneous
structure, i.e. it is homogeneously filled by a single relativistic
particle distribution, radiative and adiabatic losses alone cannot
account for the fast decrease of the emission out of the knot.  We
discuss the possible solutions to this problem and finally we describe
a scenario that seems to account for the main properties of the knot
emission.

\section{Adiabatic and radiative losses}

In this section we analyze the evolution of particles subject to both 
radiative and adiabatic cooling and we apply the results 
to the overall knot emission.

\subsection{The evolution of the electron distribution}

Let us call $N(\gamma, r)$ the particle distribution
at some distance $r$ from some initial site $r_0$, 
where $N$ describes the total number of electrons 
(not the density). 
The plasma is assumed to have bulk motion
with velocity $\beta c$ and Lorentz factor $\Gamma$.
In the absence of (re)--acceleration, $N(\gamma, r)$ is
described by (Sikora et al. 2001)
\begin{equation}
{\partial N \over \partial r} \, =\, {\partial \over \partial \gamma}
\left( N {d\gamma \over dr}\right)
\end{equation} 
This is the usual continuity equation in which time has been
substituted by the radial coordinate $r$ through
$r=\beta c\Gamma t^\prime$, and where $t^\prime$ is the
time measured in the comoving frame.
Assuming that inverse Compton losses (off CMB photons) 
dominate over synchrotron losses we have
\begin{equation}
{d\gamma\over dr} \, =\, - {1\over \beta c \Gamma} 
 \, {4\sigma_T c U_{\rm rad}\over 3 m_e c^2}\, \gamma^2
- A{\gamma \over r}
\end{equation}
where $U_{\rm rad}\, =\, 4\times 10^{-12} (1+z)^4 \Gamma^2$
erg cm$^{-3}$, and where the latter term is the adiabatic term:
$A=1$ for 3D expansion and $A=2/3$ for 2D expansion.
The solution of Eq. (2) is
\begin{equation}
\gamma\, =\, { (1-A)  \over  
\alpha\,  r+  (r_0/r)^{-A} [ (1-A)/ \gamma_0 -\alpha\, r_0 ]}
\end{equation}
where $\alpha\equiv \, 4\sigma_T U_{\rm rad}/(3 \beta \Gamma m_e c^2)$.
\begin{figure}
\psfig{figure=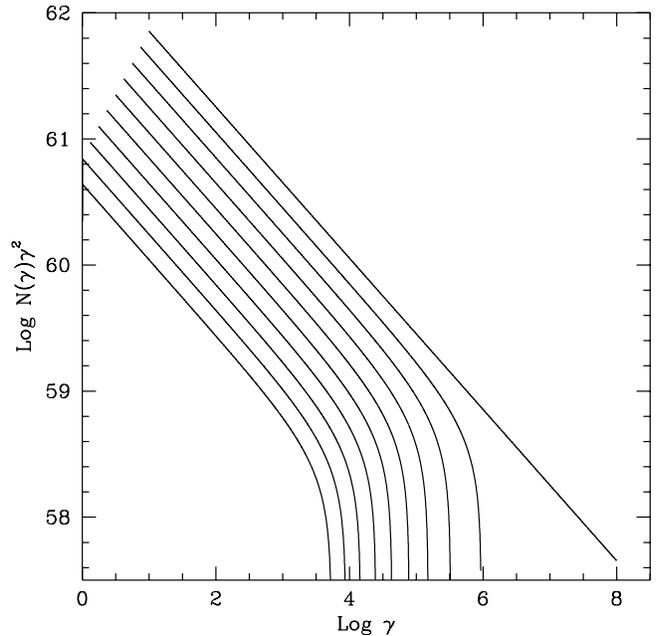,angle=0,width=9cm}
\caption{Particle distribution evolving with $r$, calculated assuming the
initial parameters as listed in Tab. 1.
Note that $\gamma_{\rm min}$ decreases with $r$ due to adiabatic losses.
To conserve the total particle number the normalization also decreases with $r$. }
\end{figure}
All particles initially between
$\gamma_0$ and $\gamma_0+d\gamma_0$, in a time $t$, (or after
some distance $r$), will form the distribution 
between the corresponding interval
$\gamma$ and $\gamma+d\gamma$, where $\gamma$ is given by Eq. (18).
We can therefore set
$N(\gamma, r) d\gamma = N_0 (\gamma_0, r_0) d\gamma_0$. From Eq. (3) we have:
\begin{equation}
{d\gamma \over d\gamma_0}  =  {(1-A)^2 \over \gamma_0^2 (r_0/r)^A } 
\left[ \alpha \, r + \left( {r \over r_0}\right)^A \left( {1-A \over \gamma_0}
-\alpha \, r_0 \right) \right]^{-2}
\end{equation}
Therefore the particle distribution at a given distance $r$ and energy $\gamma$
is given by calculating the corresponding initial $\gamma_0$
(which is transformed in $\gamma$ in the time to go from $r_0$ to $r$) and
taking into account of the corresponding differentials:
\begin{equation}
N(\gamma, r) \, =\, {N_0 (\gamma_0, r_0) \over d\gamma / d\gamma_0}
\end{equation}
(This technique has been already applied in the context of
gamma--ray bursts by Malesani 2002).
In Fig. 1 we show an example of particle evolution calculated
according to Eq. 5. 
Note that total number is conserved: since the low energy cutoff
moves to lower energies (because of adiabatic expansion losses),
the normalization decreases with $r$.

%
\begin{figure}
\psfig{figure=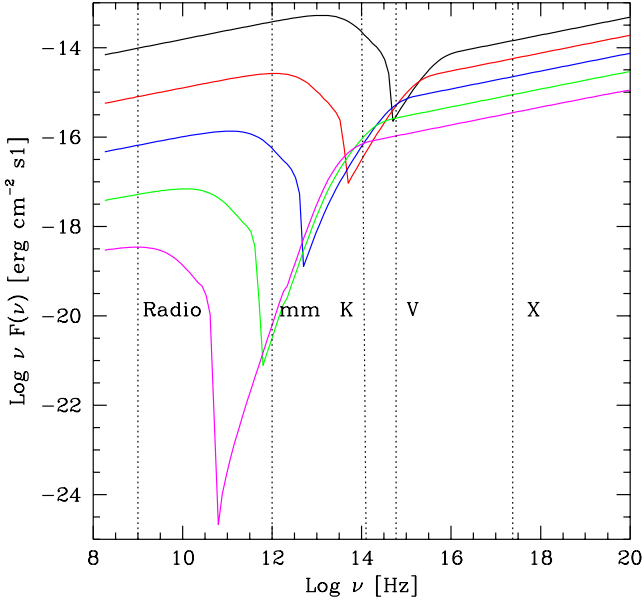,angle=0,width=9cm}
\vskip -0.5 true cm
\caption{
SEDs calculated according to the paramenters listed in Tab. 1. From top to 
bottom: $r$=20, 53, 140, 370 and 980 kpc. $\Gamma=4$ is assumed.}
\end{figure}
\begin{figure}
\psfig{figure=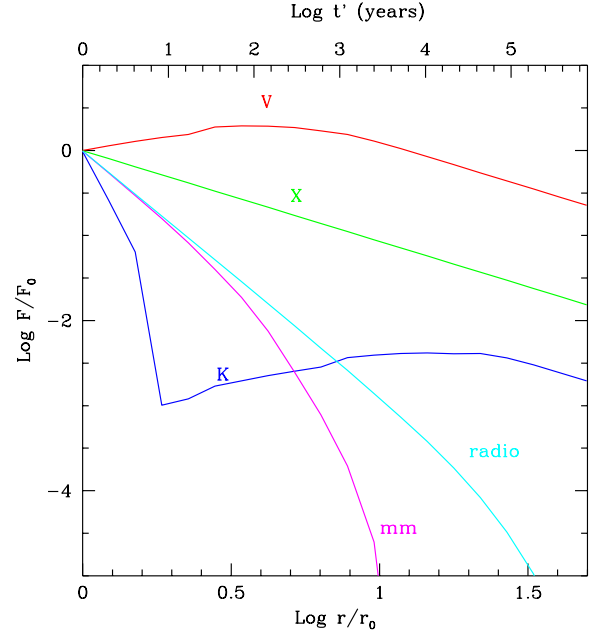,angle=0,width=8.8cm}
\vskip -0.5 true cm
\caption{Flux profiles at different frequencies (5 GHz, 0.2 mm, $K$
band, $V$ band and 1 keV), calculated assuming the parameters listed in
Table 1. $\Gamma=4$ is assumed.
This is the case of the optical being made by inverse
Compton off the CMB photons. In this case the optical {\it will
initially increase} moving away from the knot in case A (bottom
x--axis) or, alternatively (case B), it will initially increase
in time (top x--axis, time is measured in the blob's frame). See text
for details on the two descriptions.
The initial values of the fluxes at the different frequencies are: 
$F_{\rm 5~GHz}$= 209 mJy,
$F_{\rm 0.2~mm}$= 4 mJy,
$F_K= 20$ $\mu$Jy,
$F_V= 0.1$ $\mu$Jy,
$F_{\rm 1~keV}$=6.2 nJy.
}
\end{figure}
\begin{figure}
\psfig{figure=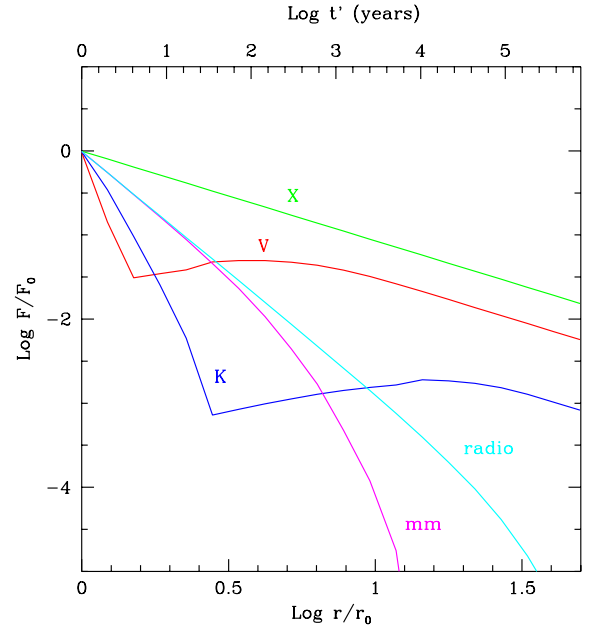,angle=0,width=8.7cm}
\vskip -0.3 true cm
\caption{Flux profiles at different frequencies, assuming the
parameters reported in Table 1. $\Gamma=4$ is assumed.
This is the case of the optical being made by synchrotron. 
The initial values of the fluxes at the different frequencies are
the same as in Fig. 3, except for the optical flux, now being $F_V$=2 $\mu$Jy.
}
\end{figure}
%


\begin{figure}
\psfig{figure=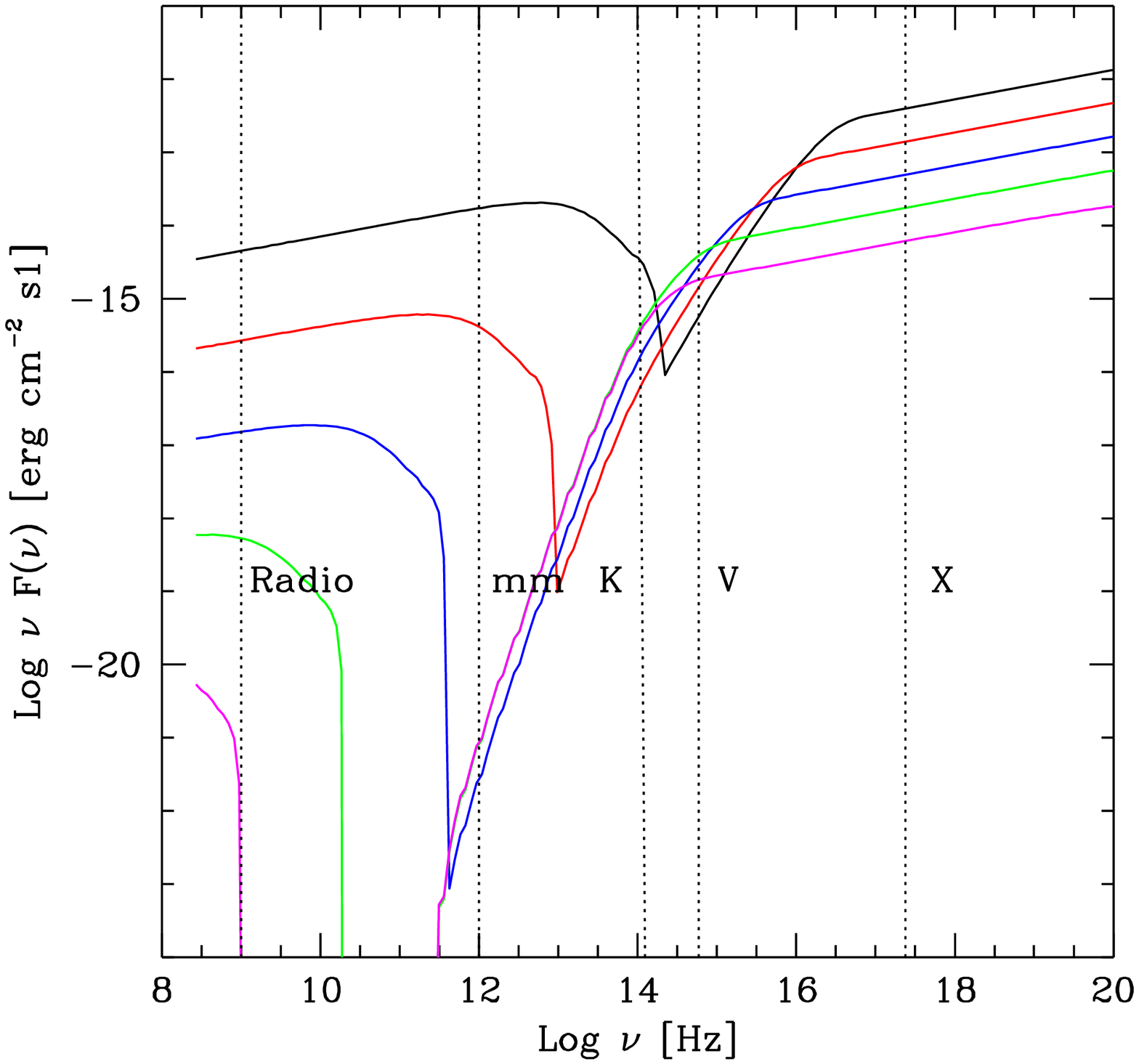,angle=0,width=9cm}
\vskip -0.4 true cm
\caption{
SEDs calculated according to the paramenters listed in Tab. 1. From top to bottom: 
$r$=20, 53, 140, 370 and 980 kpc.
$\Gamma=15$ is assumed.}
\end{figure}
\begin{figure}
\psfig{figure=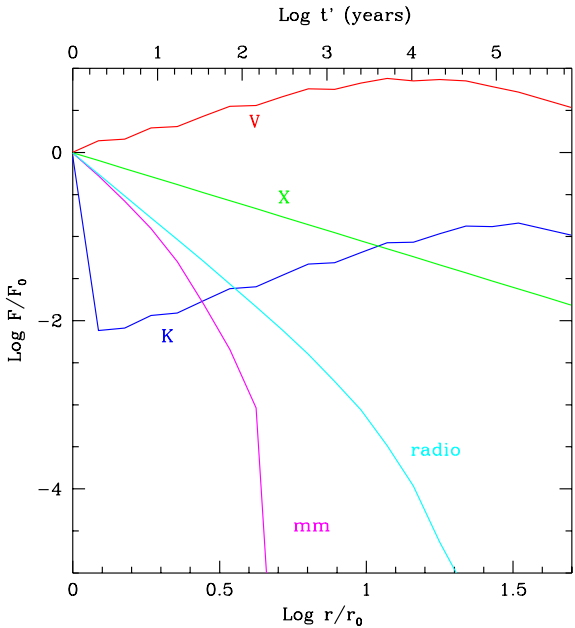,angle=0,width=8.7cm}
\vskip -0.2 true cm
\caption{
Flux profiles at different frequencies (0.2 mm, 5 GHz, $K$
band, $V$ band and 1 keV), for $\Gamma=15$, calculated assuming the parameters listed
in Tab. 1. The optical emission is due to IC off the CMB.
See that also in this case (as in Fig. 3) the optical will
initially increase moving away from the knot in case A (bottom
x--axis) or, alternatively (case B), it will initially increase
in time (top x--axis, time is measured in the blob's frame). See text
for details on the two descriptions. 
The initial values of the fluxes at the different frequencies are:
$F_{\rm 5~GHz}$= 93 mJy,
$F_{\rm 0.2~mm}$= 1.7 mJy,
$F_K= 3$ $\mu$Jy,
$F_V= 0.13$ $\mu$Jy,
$F_{\rm 1~keV}$=208 nJy.
}
\end{figure}
\begin{figure}
\psfig{figure=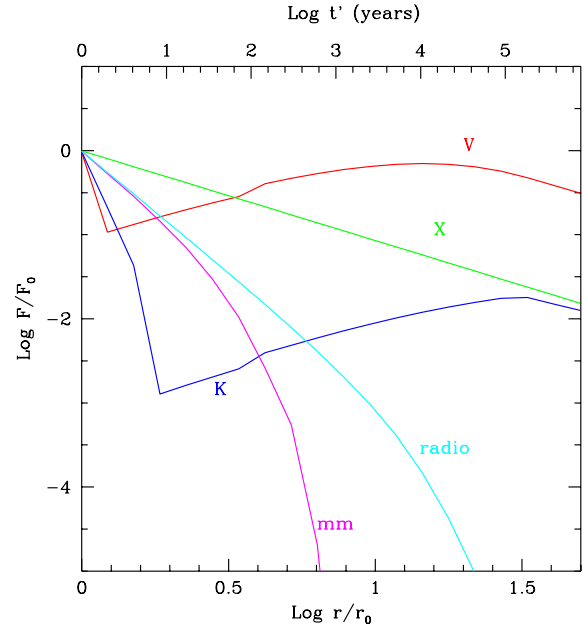,angle=0,width=8.7cm}
\vskip -0.5 true cm
\caption{
Flux profiles at different frequencies, for $\Gamma=15$, assuming the
parameters reported in Table 1.  This is the case of the optical being
made by synchrotron. 
The initial values of the fluxes at the different frequencies are
the same as in Fig. 6 except for the optical flux,
now being $F_V$=1.5 $\mu$Jy.}
\end{figure}
\subsection{Results}

Using Eq. (5) we have computed the evolution of the spectral energy
distributions (SED) as a function of $r$, assuming that $N(\gamma_0,
r_0)$ is a power law between $\gamma_{\rm min, 0}$ and $\gamma_{\rm
max, 0}$.  The initial parameters, listed in Tab. 1, have been chosen
to roughly match those found by fitting the SED of a knot of a large
scale jet visible in radio, optical and X--rays.  Note that the
magnetic field does not affect the particle evolution (since we have
assumed that the radiation energy density of the CMB dominates the
cooling), but it determines the amount of the emitted synchrotron
radiation.  We have assumed that the magnetic field scales as $r^{-1}$
(as in the case in which the magnetic field has a prevalent
toroidal component or in the case of constant magnetic flux along the
jet).  We have then assumed that the total number of electrons in a
``slice" of the jet of width $\Delta r$ is conserved, and to calculate
the emitted flux we have assumed a fixed width $\Delta r$.  We have
then consistently set the adiabatic constant $A=2/3$, corresponding to
2D expansion.

\begin{table}
\begin{center}
\begin{tabular}{|lllllllll|}
\hline & & & & & & & &\\ 
$r_0$ &$B_0$ &$A$ &$\gamma_{\rm min, 0}$ &$\gamma_{\rm max, 0}$ 
&$\Gamma$ &$\delta$ &$z$ &Fig. \\ 
kpc &$\mu$G & & & & & & & \\ 
\hline 
20   &36 &2/3 &10 &1e8   &15 &5.4 &0.72 &1    \\ 
20   &36 &2/3 &10 &2.4e5 &4  &6   &1    &2, 3 \\ 
20   &36 &2/3 &10 &5e5   &4  &6   &1    &4    \\ 
20   &10 &2/3 &10 &2.4e5 &15 &9   &1    &5, 6  \\ 
20   &10 &2/3 &10 &5e5   &15 &9   &1    &7     \\ 
0.16 &10 &1   &20 &1e6   &15 &15  &1    &8, 9  \\ 
\hline
\end{tabular}
\caption{Input parameters for the examples shown in Fig. 1--9. 
In all cases we assumed $r_{\rm max}=20 r_0$ and $N(\gamma)\propto \gamma^{-n}$
with $n=2.6$.}
\end{center}
\end{table}

Having computed the SEDs in this way, we have constructed the flux
profiles at different frequencies. 
Jets with a bulk Lorentz factor of $\Gamma=4$ and $\Gamma=15$ have been 
considered, since this is the range of values found when fitting
quasars with large scale X--ray jets 
(Tavecchio et al. 2000; Sambruna et al. 2002; Celotti et al. 2001).
Furthemore we have performed our calculations assuming that the
optical flux is produced by the initial tail of the IC process
(SEDs in Fig. 2 and Fig. 5; profiles in Fig. 3 and Fig. 6.)
or by the high energy tail of the synchrotron process
(profiles shown in Fig. 4 and Fig. 7).
The only input parameter changing between the two sets of figures
is the value of $\gamma_{\rm max,0}$. 

In Fig. 2 we show the computed SEDs (for the case of $\Gamma=4$)
assuming that the optical emission, at the starting point $r_0$, is
due to initial part of the inverse Compton scattering off the CMB
photons.  Therefore the electrons producing this radiation have a
small energy, and their cooling time is long.  In this case the
inverse Compton spectra are dominating the bolometric luminosity (this
is also true at early times, although this may not be immediately
clear from Fig. 2 which cuts the Compton spectrum before its peak),
but not by a large factor at the beginning.  At late times (or at some
distance from $r_0$) the Compton dominance increases, since the
magnetic field is assumed to decrease as $r^{-1}$, while the energy
density of the CMB is constant.  Fig. 3 shows the expected profiles of
the flux in different bands using the same parameters used to
construct the SEDs as shown in Fig. 2 and listed in Table 1.  The
x--axis of this figure can be expressed as a distance from the
starting point (bottom axis) or equivalently as the time elapsed from
the beginning of our calculations (top axis).  In the latter case the
light curves shown do not take into account light travel time effects:
since the source is very extended, we will receive light from
different evolutionary phases of the source (evolved from the near
part, younger from the far part of the source).  This is important of
course for $t<r_0/c$: at these times the observed light curve is not
the one shown in Fig. 3 (and the analogous Figs. 4, 6 and 7), but
should be calculated (as e.g. in Chiaberge \& Ghisellini 1999)
considering the different travel paths of light rays produced in
different parts of the source.  For the purposes of the present paper these
effects are not crucial, but they will be properly considered in our
future studies.  Note that if the profiles (i.e. bottom x--axis) shown
in Fig. 3 correspond to emission from a standing shock (see below),
and so a quasi--stationary case, then light travel time effects are not
important, and Fig. 3 (and the other analogous figures) already fully
describes the expected behavior.

Fig. 4 shows the profiles assuming that the optical emission is
due to the high--energy tail of the synchrotron component.
The only parameter different from those adopted for Fig. 2 and Fig. 3 is
$\gamma_{\rm max, 0}$, which is now slightly smaller (see Tab. 1).

Fig. 5 shows, by contrast, the SEDs predicted in the case of
$\Gamma=15$.  We have assumed that in this case the Compton emission
is more dominant than in the previous case, due to the increased
energy density of the CMB as seen in the comoving frame ($\propto
\Gamma^2$).  The other input parameters are listed in Table 1.  Note
that since the comoving CMB energy density is larger than in the
previous case, the corresponding radiative cooling times are shorter,
making the evolution of the high energy electrons faster, as can be
seen comparing the radio and far IR spectra in Fig. 5 and Fig. 2 for
late times (or for large distances).  As in Fig. 2, the SEDs shown in
Fig. 5 correspond to the case of the optical emission produced by the
initial part of the inverse Compton radiation.  Fig. 6 shows the
corresponding profiles of the flux at different frequencies, while
Fig. 7 shows the profiles assuming that, at the start, the optical
flux is due to the tail of the synchrotron emission.  Fig. 5 shows
also that the synchrotron and IC emission components becomes
``detached" at late times, when only the low energy electrons survive.

On inspecting the figures showing the flux profiles, a general trend
emerges: the synchrotron produced fluxes are dimming faster than the
IC/CMB ones, due to i) the different electron energies involved and
ii) the decreasing magnetic field.  In all cases the X--ray fluxes are
roughly linearly decreasing with the distance $r$, while the initial
decrease of the fluxes produced by the synchrotron process (away from
the cutoffs) scale roughly as $r^{-3}$.  As discussed below, this
different behavior will be crucial when comparing our results with
already existing data.

As mentioned in the Introduction, there are two possible cases
for interpreting the results shown in Figs. 3, 4 and in Figs. 6, 7. 
Let us call them:

{\bf case A}: streaming of particles from a stationary (in the observer
frame) shock (bottom x--axis, distance from the acceleration site);

{\bf case B}: rigidly moving blob whose internal particles have been
accelerated at $r_0$ (top x--axis, time from the acceleration event in
the blob reference frame).

If we are in case A (standing shock), then Figs. 3, 4, 6 and 7 
directly show the
profiles that would be observed in deep images of the jet for $r>r_0$
(assuming that the standing shock lives for a sufficiently
long time to allow particles to move from $r_0$ to $r$). In the case
of Fig. 3 and Fig. 6 for which the optical flux is due to the low--energy tail of
the IC component, the optical flux {\it increases} with $r$.
This happens despite the decrease in the normalization of $N(\gamma)$,
and it is due to the fact that the optical is produced by the initial
tail of the Compton emission (steeply rising with frequency), which 
increases its flux because of the decrease of $\gamma_{\rm min}$.
In other words, in this case the overall spectrum shifts to lower 
frequencies and to lower fluxes: in the power law part of the spectrum
the flux density decreases, while at the low energy end the flux density
increases.
In fact, when the optical flux is due to the power law part of the 
external Compton spectrum (for $r/r_0>10$), it decreases as the X--ray flux. 

In case B (evolving blob), the particle distribution at a given $r$ is
not directly related to the bright knots we are detecting, but is
instead the ``fossil" distribution of other unknown, acceleration
phases occurred in the past (at $t^\prime=t^\prime_0$ on the top
x--axis).  Assuming then that there have been such phases in the past,
with the optical flux initially produced through IC (Fig. 3 and
Fig. 6), we should observe some bright optical knots with relatively
weak X--ray and radio fluxes (with respect to the HST/Chandra bright
knots).  Knots which are bright only in the optical band have not been
detected, and this could lead to the conclusion that the optical
emission is always due to the high energy tail of the synchrotron
spectra, and therefore to very high energy electrons.  However, as
will be discussed below, this conclusion is premature, since there may
be an alternative possibility, in which the knot emission is
considered as the result of the contribution of many sub--knot
emission sites.

\section{Comparison with observations}

The results of the analysis described above can be compared with real
profiles of nearby jets.  One of the sources with the best published
data is the luminous jet of 3C 273 (Sambruna et al. 2001; Marshall et
al. 2001)

The jet presents numerous knots visible in radio, optical and X--rays. 
Sambruna et al. (2001) and Marshall et al. (2001) reach different
conclusions about the viability of a synchrotron interpretation for
the X--ray emission in knots A and B1 (the two most luminous portions
of the jet).  This is due to the different slopes of the optical
spectrum derived by the two groups: Sambruna et al. (2001) present an
optical spectrum with a slope systematically steeper than that of
Marshall et al. (2001). For this reason Sambruna et al.  can easily
rule out the possibility that a unique spectral component from radio
to X--rays can account from the overall spectrum, while the opposite
conclusion is preferred by Marshall et al.

Despite this disagreement, both groups agree on the fact that the
optical flux is due to synchrotron, and on the fact that maps at
different frequencies yield knots of the same dimensions, and similar
flux profiles.  In other words, there appears to be no
frequency--dependence of the size of knots.  

However, according to the analysis given above, we would expect to
find the radio knots to be larger than the optical ones, due to the
longer lifetime of the radio-emitting electrons. An even larger
region is expected in the case of X--ray knots produced by the IC
process (i.e. by very low--energy electrons).  Contrary to this
expectation, it appears that knots have almost the same size at
different frequencies.

In the following we examine what at first sight can be the possible
and most plausible solutions:

\begin{itemize}

\item
Consider a {\bf small and single standing shock}.  Suppose that the
acceleration site is very compact, much smaller than the cross
sectional radius of the jet, i.e. smaller than one arcsec.  The
angular resolution of Chandra in this case is not enough to resolve it
in X--rays.  In this way the adiabatic losses are much more effective.
They corresponds to take, as $r_0$ in our figure, a region which is
much less than a kpc (say a factor 10 less).  In this case there are
enough ``doubling radii" to let adiabatic losses to decrease the flux
(as in our figures).  {\it But the angular resolution of HST allows to
resolve the emission region in the optical, and show that it cannot be
smaller than about 1 arcsec.}  This implies that the acceleration site
is extended (at least 1 arcsec), or made by several sub--units,
contrary to our assumption.

\item
Consider {\bf moving blobs}.
In this case we see a snapshot of the blobs which are now active, and moving.
The emission at different frequencies can in this case be cospatial, with no emission
between the knots.
The knots are observed to be active simultaneously according to our (observing) time, 
implying that they have been lighted up at different intrinsic times
(because of light travel time effects).
If these knots were the only existing ones, this requires a strong fine tuning. 
On the other hand, if the jet has many knots that randomly (in time and in location)
light up, we always have the possibility to observe a few of them ``on"
(thus avoiding the fine tuning problem), but automatically this requires the 
existence of many ``fossils" of previous active phases,
emitting some X--rays, some radio, and no optical [i.e. with only the low
energy part of $N(\gamma)$ surviving].
Since we observe no ``fossil" knot, we disfavor this possibility. 
Yet another alternative is to assume that the jet
really has only a few knots, which are {\it always} active.
In this case we do not observe fossils, but we require that the
particles in each knot are continuously reaccelerated.
However this leads to another problem: in fact if the same particles
are continuously reaccelerated, they should reach a typical energy 
(at which the cooling and acceleration  timescales are equal), 
causing a pile--up at some energy, which is not observed.
A possible way out is to assume that particles are only episodically 
reaccelerated, but this would require a strong fine--tuning between the duty 
cycle of the acceleration episodes and the cooling time of the optical 
electrons (the two times must be comparable).

\end{itemize}

We conclude that simple alternatives to the basic scenario (basic in the sense
of a single homogeneous region emitting by synchrotron and external Compton)
do not work or require fine--tuning.
We are then forced to explore a slightly more complex alternative.
In the next section we will then consider the possibility of knots composed
of several sub--units (which we may identify as the acceleration sites), 
in order to make expansion losses more effective.

\section{The clumping scenario}

Assume that the knot is not homogenous, but is instead  
formed by several smaller substructures, i.e. several clumps. 
The key point of this idea is that small clumps can expand much more than 
a unique knot and therefore adiabatic losses can play a more important 
role than in the homogeneous case discussed above.

\begin{figure}
\psfig{figure=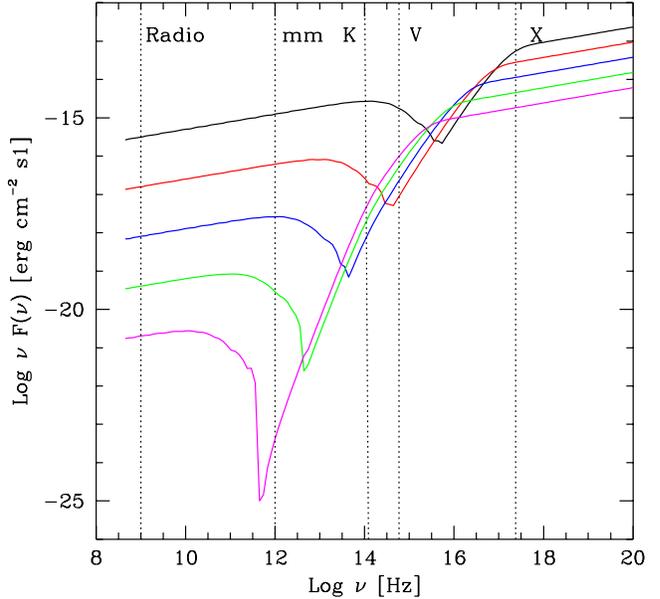,angle=0,width=9cm}
\vskip -0.4 true cm
\caption{
SEDs calculated in the ``mini--knots" scenario, using the paramenters listed in Tab. 1.
This illustrates the spectral evolution of a single mini--knot, 
for different expansion stages. From top to bottom: $r=$0.16, 0.28, 0.5, 0.9 and 1.6 kpc.
Note that for this case we assumed a 3D expansion, and correspondingly used $A=1$. }
\end{figure}
\begin{figure}
\psfig{figure=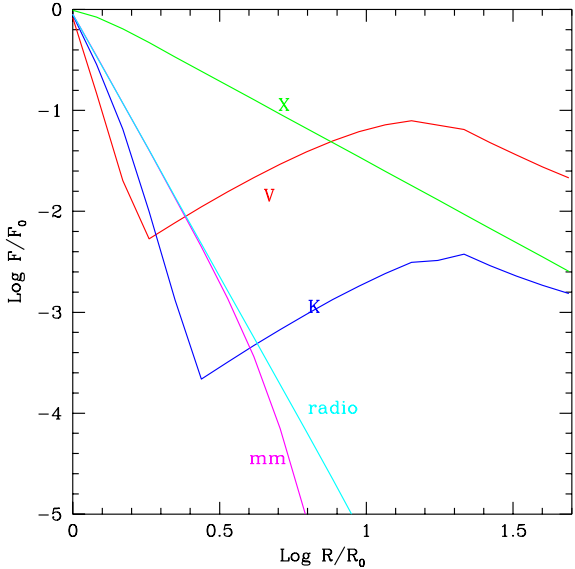,angle=0,width=8.7cm}
\vskip -0.2 true cm
\caption{
Flux profiles at different frequencies (0.2 mm, 5 GHz, $K$
band, $V$ band and 1 keV), for $\Gamma=15$, in the case of a single
mini--knot, using the same parameters as in Fig. 8.}
\end{figure}

Assume then several acceleration sites which are much smaller than the
transverse dimension of the jet.  Let us call them ``mini--knots".
Assume a number $N$ of mini--knots, and for simplicity assume that all
of them are the same, with the same density $n_i$, radius $R_i$, volume
$V_i$ and bulk Lorentz factor $\Gamma_i$.  The filling factor of the
mini--knots is $f\equiv NV_i/V$ where $V$ is the volume of the large
knot.  $R_i$ is limited to be smaller than -- say -- one tenth of the
total size of the knot (to have enough expansion losses within the big
knot).  With these assumptions also the magnetic field $B_i$ must be
the same, whatever the radius and number of mini--knots.  This
follows from the fact that in each mini--knot the magnetic field to
CMB energy density ratio must be equal to the one-zone case, since the
resulting spectrum must have the same synchrotron to IC luminosity
ratio. Since $U_{\rm rad}$ in the mini-knot scenario is equal to the
the one-zone scenario (as long as $\Gamma _i$ is the same), also the
magnetic field must assume the same value in both scenarios.

Just for illustrative purposes, let us compare the case of a single
homogeneous model in which the size of the emitting region is the
entire big knot (we may call it the one--zone case: this is the case
discussed up to now in the literature) with the clumping case.  Assume
for simplicity that the total number of emitting particles is the same
in the two cases, as well as the magnetic field and the bulk Lorentz
factor.  What varies is the density and the size of the mini--knots:
$n_i$ must be a factor $f$ larger than in the one--zone case.  With
these assumptions, the average bulk kinetic power (dominated by the
cold proton component) in the case of the mini--knots is the same as in the
one--zone case: the total number of particles is the same, and the
time needed for them to cross a section of the jet is the same in both
scenarios.

There is no obvious lower limit to the size $R_i$ of the mini--knots,
but it must be noted that (within our assumptions) for a fixed number
$N$ of mini--knots, the smaller their size, the larger their density,
and the larger the deviation from equipartition between particle and
magnetic energy densities.  This deviation scales as the ratio of the
particle density in each mini--knot and the density in the one--zone
case, and therefore it scales as $f$.  Therefore it is possible to
have many (and small) mini--knots with a small filling factor, but
this implies a large deviation from equipartition.  


With respect to the one--zone case, each mini--knot has an enhanced 
particle pressure, and it is thus likely that it is overpressured
with respect to its surroundings, implying expansion.

In conclusion, we propose that there are several mini--knots within
the larger blob, which are the sites of particle
acceleration, which are denser and hotter than the surroundings,
leading automatically to the expansion of the entire mini--knot.
What we see is then the convolution of the radiation produced by 
several expanding mini--knots.

The formation mechanisms of these ``mini--knots" is outside the main
goal of this paper; here we would only like to mention possible
instabilities, clouds (filaments) crossing the jet (as proposed by,
e.g., Blandford \& Konigl 1979, although they considered this process
at much smaller scales), or instabilities triggered by entrained
gas that can act as an obstacle for the flow, forming mini--shocks.
Another possibility is that particle acceleration is induced by
reconnection events in the jet (e.g. Drenkhan \& Spruit 2002). In this
case the mini--knots could be identified as the reconnection sites.

If such mini--knots exist, then the problem to let low energy
electrons adiabatically cool before they reach the borders of the
bigger knot as detected by Chandra is solved.  
To illustrate this, we show in Fig. 8 the SEDs corresponding to 
a single mini--knot and in Fig. 9 the corresponding flux profiles 
at different frequencies.
Since $r_0$ is much smaller than in the previous cases, all
the timescales are shorter, changing the relative importance 
of the radiative and adiabatic losses.
Note that, in Fig. 9, the entire
range of the x--axis is now 1 arcsec, as opposed to 10 arcsec of the
previous ones (for a source at $z\sim 1$).

While cooling for radiative and adiabatic losses, and while the
mini--knot expands, the electrons emit some radiation even outside the
acceleration sites.  High energy particles, producing the optical
synchrotron radiation, are the ones cooling the most, making the
optical flux to decrease at the faster rate initially.  Then, when
$\gamma_{\rm min}$ is low enough to produce optical radiation by the
inverse Compton process with the CMB photons, we have an increasing
optical flux (until it is produced by the rising initial tail of the
IC spectrum).  The same situation occurs (but later) for the infrared
flux.  The X--ray flux instead steadily decreases, being always
produced by low energy electrons.  In conclusion, having enough
angular resolution, one should resolve each mini--knot, especially in
the optical and infrared bands, where the contrast between the
emission inside and outside the mini--knots is the largest.  However,
the flux between the knots never vanishes, and remains at a
considerable level especially in the X--ray band.

Another point that must be considered is the nature of the medium
 between the ``mini-knots''. If particle acceleration is concentrated
 in the ``mini-knots'', the possibility that appears the most natural
 is that the medium is mainly composed by {\it cold} particles,
 perhaps embedded in a very low magnetic field.  These particles would
 emit through IC/CMB: the emission is concentrated in the IR-optical
 regime, at $\sim 10^{14}$ Hz. In this case there would be the
 possibility that this material is detectable in a near future with
 NGST, apearing as a ``diluting'' emission between mini-knots. The
 possibility to detect the emission mainly depends on the density of
 the medium: even an upper limit could help to better characterize the
 density contrast between the clumps and the external medium.

\section{Summary}

We have discussed how the presence of low energy electrons in jets,
predicted by the IC/CMB model, poses important problems related to
their extremely long radiative cooling time.  In fact we know that the
X--ray emission (as well as the optical and radio emission) comes in
localized regions of the jet (the knots), despite the fact that the
X--ray emitting electrons cannot radiatively cool inside these small
regions.  This motivated the exploration of the possible effects of
adiabatic losses that for low energy electrons are much more important
than radiative losses.  We have shown the predicted profiles at
different frequencies under different assumptions.

In particular we have identified two possible cases: in case A we
considered a standing (in the observer frame) shock, while in case B
the emission we observe is produced in a moving blob.  Some general
consequences appropriate for future and more detailed observations
will be briefly discussed below and in a forthcoming paper.

However, the existing observations already pose a problem.  In fact
comparing our results with the data of nearby aligned radio sources
(i.e. nearby blazars, such as 3C 273), we found that in any case (A or
B) radiative and adiabatic losses cannot explain what is observed.  In
case A we would expect radio (and perhaps X--ray) knots larger than
the optical ones, contrary to what is observed.  In case B we would
expect a number of unobserved ``fossil knots", bright in the radio and
especially in the X--ray bands, but much fainter in the optical.

A possible way out of this puzzle is to admit that the emitting region
is clumped in several mini--knots, allowing particles to cool by
adiabatic expansion much more than in the one--zone case.  These
clumps can originate from different processes.  An important
consequence of this scenario is that, since the magnetic field
energy density has to be the same as in the case of the one-zone
scenario, the emitting region is far from equipartition, with the
particle energy density dominating over the magnetic one.

\section{Discussion}

In large scale jets, with moderate values of the energy densities
involved (both magnetic and radiative), the radiative cooling times
are very long for all but the TeV electrons responsible for the high
energy tail of the synchrotron emission.  In particular, in the IC/CMB
models, the X--ray flux is produced by electrons with random Lorentz
factors of the order of 10--100, implying that they never cool through
radiation losses.  Adiabatic losses must then play a crucial role to
explain the knotty morphology of large scale X--ray/optical/radio
jets.

What we see can be the result of particle acceleration occurring in
two different scenarios: acceleration can occur either at the front of
a standing shock or accelerated particles can be injected throughout a
blob moving along with the jet.

In the first case we expect to observe an extension of the emission
region downstream to the shock (at larger distances from the shock
front), whose size depends on the observing frequency.  In particular,
since X--rays are produced by low-energy electrons, the size of the
X--ray emitting region must be larger than the radio one.  The optical
behavior is more complex, being due initially to the highest energy
and rapidly cooling electrons, and later being due to the lowest
energy electrons through IC.

In the moving blob case, particles are confined and evolve within the
blob, and the particle evolution translates into a time dependent
appearance of the blob itself.  In particular, also old, ``fossil"
blobs are expected to be observable, bright especially in the X--ray
band, relatively less in the radio and optical.  Even taking into
account adiabatic losses, the lifetime of the X--ray emission is long
(compared to $R/c$, where $R$ is the size of the blob), and therefore
the number of these ``fossil blobs" should be larger than the young
ones.

These considerations point towards the possibility to discriminate
between the two scenarios: observing "X--ray trails" more elongated
than the radio and the optical ones would favor the standing shock
scenario, while observing isolated X--ray knots (with weaker radio and
optical emission, but cospatial) would favor the moving blob scenario.
In order to perform such discriminating observations one needs
sub--arcsec angular resolution at flux levels of the order of
$10^{-15}$ erg cm$^{-2}$ s$^{-1}$ or better, at least in the
optical/IR, possibly in the mm and radio domains.  In general, NGST,
LBT and ALMA are then required.

Nearby blazars can already give some indications, since for them HST
already can probe the interesting spatial scales.  We have compared
our results with the existing data of 3C 273.  What we found is at
first sight puzzling, since the flux outside the knots decreases fast
and no ``fossils" are observed.  We have then proposed that what
appears a single knot is instead the convolution of several smaller
emission sites.  This allows particles to adiabatically cool within
the observed knot, allowing the flux to decrease fast outside it.

A natural consequence of this clumping scenario is that the emission
can be variable on short (months--years) timescales.  We note that,
very recently, the knot HST--1 in M87 (emitting by synchrotron from
the radio to the X--ray band), has been claimed to vary on a timescale
of months in the X--ray and optical bands, and not in the radio
(Harris et al. 2002).  These variations have been attributed to the
radiative cooling of the synchrotron emitting electrons.  However, the
dimension of knot HST--1 is of the order of $\sim 100$ pc, implying a
minimum observable variability timescale of $\sim$300 years,
corresponding to the light crossing time.  If the HST--1 blob is
instead composed of several mini--knots, the minimum variability
timescale becomes the light crossing time of each single mini--knot
(as long as the total number of mini--knots is not large, otherwise
the amplitude of variability is diluted and becomes unobservable).
Also in M87' -- suggest rewriting this sentence as follows `M87,
then, also shows some evidence for clumped substructure in the jet
knots, as we have suggested in the case of 3C273 using a completely
independent argument.

\begin{acknowledgements}
We thank the Italian MIUR and ASI for financial support.
We thank T. Belloni and D. Malesani for useful discussions.
\end{acknowledgements}

\end{document}